\documentclass[aps, pra, reprint, showpacs]{revtex4-1}
\usepackage{graphicx}
\usepackage{amsmath}
\usepackage{amssymb}
\usepackage[colorlinks, allcolors=blue]{hyperref}
\usepackage{hypcap}
\newcommand{\pref}[2]{\hyperref[#1]{\ref{#1}(#2)}}
\newcommand{\eqpref}[1]{\hyperref[#1]{(\ref{#1})}}
\newcommand{\wt}[1]{%
  \mspace{2mu}%
  \tilde{\mspace{-2mu}\rule{0pt}{1.3ex}\smash[t]{#1}}%
}
\begin{document}
\title{Atom-optics simulator of lattice transport phenomena}
\author{Eric J. Meier}
\author{Fangzhao Alex An}
\author{Bryce Gadway}
\email{bgadway@illinois.edu}
\affiliation{Department of Physics, University of Illinois at Urbana-Champaign, Urbana, Illinois 61801-3080, USA}
\date{\today}
\begin{abstract}
We experimentally investigate a scheme for studying lattice transport phenomena, based on the controlled momentum-space dynamics of ultracold atomic matter waves. In the effective tight-binding models that can be simulated, we demonstrate that this technique allows for a local and time-dependent control over all system parameters, and additionally allows for single-site resolved detection of atomic populations. We demonstrate full control over site-to-site off-diagonal tunneling elements (amplitude and phase) and diagonal site-energies, through the observation of continuous-time quantum walks, Bloch oscillations, and negative tunneling. These capabilities open up new prospects in the experimental study of disordered and topological systems.
\end{abstract}

\maketitle
In recent years, owing to their intrinsic purity and the high degree of control available in them, atomic, molecular, and optical systems have found widespread use in the simulation and study of condensed matter physics phenomena~\cite{Bloch-RMP08,Lewenstein-Mimicking-,Chien2015}. In particular, ultracold atoms in pristine optical lattices play host to a number of textbook phenomena typically associated with electrons in crystalline solids, including Bloch oscillations in electric fields~\cite{BenDahan-Bloch}, Dirac-like dispersion in graphene lattices~\cite{Tarruell-Dirac-2012}, and the formation of correlated Mott insulators due to interactions~\cite{Greiner-SF-MI-2002}. Ever more rich and complex optical potentials, relevant e.g. for the study of topological materials, are now being formed by the controlled superposition of multiple lattice structures~\cite{Salomon-projective-1998,Sebby-superlattice,Folling2ndOrder,Becker-triang,Tarruell-Dirac-2012,Jo-Kagome}. An alternative approach for the engineering of lattice structures has been taken in photonic platforms used for the simulation of particle transport~\cite{SzameitReview-2010,Quasi1D-2D-Kraus-2012A,Segev-LightLocal-2013,Rech-TopFloquet-2013,Hafezi-ImagingTop-2013}, based on the microscopic control over individual lattice sites and tunneling links.

Here, we experimentally realize a recently proposed technique~\cite{Gadway-KSPACE} that allows for a similar level of microscopic control to be exacted in an ultracold atomic physics setting. This atom optics-based~\cite{Adams-94,Berman-B97,Cronin-RMP} approach uses stimulated Bragg transitions to control the coherent coupling between a large number of plane-wave momentum states of a Bose--Einstein condensate, mimicking coherent particle tunneling between sites in a lattice array. This builds on a large body of work using atomic momentum-space dynamics for the study of quantum transport phenomena~\cite{Moore-Qdeltakicked-1995,Hensinger-Phillips-DynTunnel,Steck-ChaosTunnel,Ryu-HighOrderRes-2006,Chabe-AndersonMetal-2008,Lemarie-CriticalStateAnderson-2010,Lopez-Universal,Gadway-rotors}, introducing the ability to locally control many momentum-space couplings.

We demonstrate the ability to construct tight-binding models for the simulation of coherent particle transport, with individual control over all nearest-neighbor tunneling amplitudes and phases as well as individual site energies. We illustrate this control by studying continuous-time quantum walks with hard-wall system boundaries, Bloch oscillations in the presence of an effective electric field, and time-dependent control over tunneling phases. We anticipate that these unique capabilities will pave the way for novel studies of transport in disordered and topological systems.

\emph{Building a momentum-space lattice.}~The scheme that we explore has been described in full detail in Ref.~\cite{Gadway-KSPACE}, and we review it briefly~\cite{2015Meier-footnote-SuppMat}. We use the controlled evolution of momentum-space distributions of cold atomic gases to emulate the physics of single electron transport in tight-binding lattice models. Controlled coupling between discrete free-particle momentum states is achieved through stimulated two-photon Bragg transitions~\cite{Kozuma-Bragg,Denschlag-02}, driven by counterpropagating laser fields detuned far from atomic resonance. The lattice laser wavevector $k$ determines a discrete set of states $\psi_n$ with momenta $p_n = 2n\hbar k$ that may be populated from an at-rest condensate through the stimulated exchange of photons between the laser fields. The quadratic energy-momentum dispersion of the massive atoms defines a unique energy difference and Bragg transition frequency $\omega^{\text{res}}_n = (2n+1)4 E_R/\hbar$, with $E_R = \hbar^2 k^2 / 2M$ the recoil energy and $M$ the atomic mass, for each pair of neighboring states $\psi_n$ and $\psi_{n+1}$.

By writing multiple frequency tones -- controlled in amplitude, frequency, and phase -- onto one of the interfering laser fields, multiple two-photon Bragg transitions are simultaneously driven with spectrally resolved control at the level of individual Bragg links. We are able to realize highly tunable single-particle Hamiltonians of the form
\begin{equation}
\hat{H}_{\mathrm{eff}} \approx \sum_n [ \varepsilon_n |\wt{\psi}_n \rangle \langle \wt{\psi}_n | + t_n ( e^{i\varphi_n} |\wt{\psi}_{n+1} \rangle \langle \wt{\psi}_n |  + \mathrm{h.c.}) ] \ ,
\label{EQ:e0b}
\end{equation}
with arbitrary control over all effective tunneling amplitudes $t_n$, tunneling phases $\varphi_n$, and site energies $\varepsilon_n$ enabled in a local fashion through control of the multi-frequency global addressing field. The tunneling amplitudes are controlled through the two-photon Rabi coupling strengths of different Bragg transitions, the tunneling phases through the phase differences of the fields driving these Bragg transitions, and the site energies through detunings from the Bragg resonance conditions, with the states $|\wt{\psi}_n \rangle$ related to the bare momentum states $|\psi_n \rangle$ by a simple local transformation~\cite{2015Meier-footnote-SuppMat}. These parameters may additionally be made time-dependent through dynamical variation of the multi-frequency laser field. While the control of tunneling phases and site energies are essentially exact (through direct digital radiofrequency engineering), the tunneling amplitudes are sensitive to the alignment of laser beams with respect to our atomic samples, and are subject to typical uncertainties on the order of a few percent.

\emph{Experimental apparatus.} Our experiments are based on the controlled laser addressing of pure Bose-Einstein condensates of roughly \textbf{$5 \times 10^4$} $^{87}$Rb atoms, produced via gravitationally assisted evaporation in an optical dipole trap~\cite{2015Meier-footnote-SuppMat}. Following adiabatic decompression of the trap stiffness~\cite{Kozuma-Bragg}, the condensate is largely confined by a single far-detuned laser beam (OT1) oriented in the horizontal plane, having wavelength $\lambda = 1064$~nm and a beam waist ($1/e^2$ radius) of $\sim80~\mathrm{\mu m}$, such that it is spatially extended along one direction. Weak additional confinement is provided by two other crossed laser beams (wavelengths 1070~nm, beam waists $\sim250~\mathrm{\mu m}$), resulting in trapping frequencies of $\omega_{\{1,2,3\}} \sim 2 \pi \times \{130,10,130\}$~Hz.

\begin{figure}[t]
\capstart
\vspace{2pt}
\includegraphics[width=\columnwidth]{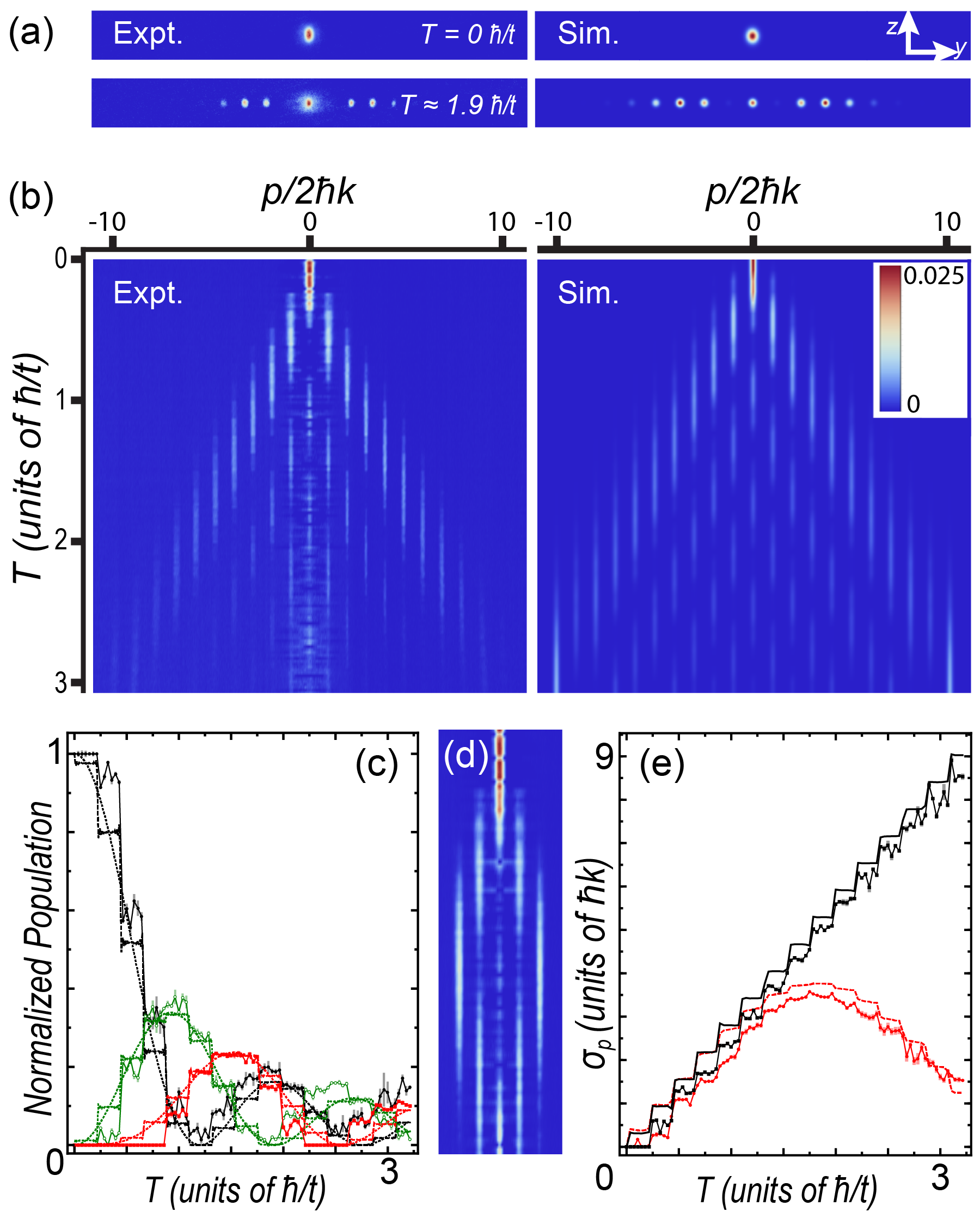}
\caption{Continuous-time quantum walks on unbounded and bounded lattices.
(a)~Experimental and simulated time-of-flight absorption images of atomic populations in momentum orders $\psi_n$ (with momenta $p_n = 2\hbar k$), before laser addressing and after $500~\mathrm{\mu s}$ of evolution with uniform nearest-neighbor couplings $t_n = t \approx 0.3 E_R$ and on site energies $\varepsilon_{n+1}-\varepsilon_n=0$ (for $n \in \{-10,9\}$). The typical uncertainty in $t_n$ is on the few percent level.
(b)~Integrated (along $z$) momentum spectra, normalized with respect to the total atom number (color scale at right), are plotted as a function of evolution time $T$ (with units $\hbar / t$).
(c)~Normalized populations of the zeroth, first, and second (black circles, open green circles, and red squares) momentum orders vs. evolution time. The dotted curves are solutions to the effective dynamics described by Eq.~\eqpref{EQ:e0b}, while the dashed curves include off-resonant Bragg coupling terms, as described in the text.
(d)~Experimental integrated momentum spectra vs. evolution time, with uniform $t_n$ truncated such that only five sites are coupled ($t_n = t \approx 0.3 E_R$ for $n \in \{ -2,1\}$).
(e)~Standard deviation of the population distributions for an unbounded lattice (black squares, solid theory line) and a bounded lattice (red circles, dashed theory line).
}
\label{FIG:fig1}
\end{figure}

The trapping laser beam OT1 serves as one of the two interfering fields driving momentum-space dynamics. A counterpropagating laser beam, composed of multiple frequency components, is derived from OT1 through diffraction from a pair of acousto-optic modulators driven by controlled multi frequency rf signals, and aligned so that it counterpropagates with respect to OT1~\cite{2015Meier-footnote-SuppMat}. The interaction of the atoms with these two interfering laser beams, with wave vectors $k = 2\pi/\lambda$, results in driven momentum-space dynamics characterized by a recoil energy $E_R \approx h \times 2.03$~kHz and changes in velocity by $\pm 2 \hbar k / M \approx \pm 8.6~\mathrm{\mu m/ms}$. After a chosen evolution time, during which the momentum-space dynamics are governed by the effective Hamiltonian of Eq.~\eqpref{EQ:e0b}, all laser fields are extinguished and the atoms are allowed to freely fall and expand in time of flight (TOF). An absorption image is taken after 18~ms of TOF, revealing the populations in the different momentum orders (states $\psi_n$).

\emph{Quantum walks on uniform lattices.} We begin our experimental exploration of this scheme by studying one of the simplest scenarios of particle transport on a 21 ``site'' lattice with uniform tunneling energies ($t_n = t \approx 0.3 E_{R}$) and site energies ($\varepsilon_{n+1} - \varepsilon_{n} = 0$). Here we observe continuous-time quantum walks (CTQWs) with a collection of atoms, which exhibit many of the hallmark features displayed by discrete-time quantum walks \cite{Aharnov93}, as realized in experiments with single atoms \cite{Karski174}, ions \cite{ZahringerIon}, and photons \cite{Schreiber10,Broome10}. Starting from an initially localized state, this textbook situation of quantum walking leads to wave packet spreading characterized by an increase of the momentum-space standard deviation as $\sigma_{p}(T) = \alpha T^\beta$, where $T$ is the evolution time. The CTQWs that we explore are characterized by ballistic wave packet spreading ($\beta = 1$), whereas classical random walking would result in diffusive spreading ($\beta = 1/2$). We expect that in our case of uniform tunneling energy $t$ the constant $\alpha$ is given simply by $\sqrt{8}kt$. Shown in Figs.~\pref{FIG:fig1}{a}-\pref{FIG:fig1}{c} are experimental data of the CTQW of our atoms in momentum space, along with numerical simulations. Ballistic spreading can be clearly observed in the momentum distributions in Fig.~\pref{FIG:fig1}{b} and in the evolution of $\sigma_p$ in Fig.~\pref{FIG:fig1}{e}(black line).

We note one interesting aspect of the population dynamics~\cite{2015Meier-footnote-SuppMat} shown in Fig.~\pref{FIG:fig1}{b} that is not captured by Eq.~\eqpref{EQ:e0b}: the appearance of small steplike jumps in the populations as opposed to perfectly smooth variations with time. The effective dynamics governed by Eq.~\eqpref{EQ:e0b} emerge in the limit where the energy scale set by the spacing between the unique Bragg resonances, i.e. $\hbar \delta \omega^{\text{res}} = 8 E_R$, far exceeds all other energy scales -- specifically all tunneling energies (two-photon Rabi coupling strengths) and all differences between adjacent site energies (detunings from Bragg resonance). This small steplike behavior comes about because, while only one particular frequency component of the interfering fields contributes resonantly to a given Bragg transition, all of the other components contribute in a small, off-resonant fashion.

The ability to study transport dynamics with local detection is one powerful capability of the explored scheme. Moreover, this scheme allows for local control over parameters of the simulated hopping model. To demonstrate this capability, we now report on dynamics in a system where we impose hard-wall-like open boundary conditions, simply by truncating the spectrum of applied frequency components such that a limited number of momentum states are resonantly coupled (in this case  $n \in \{-2,1\}$). These dynamics, showing ballistic spreading at short times followed by reflection from the open boundaries, are shown in Fig.~\pref{FIG:fig1}{d}, with a comparison of the evolution of $\sigma_p$ to the ``unbounded'' case ($n\in\{-10,9\}$) in Fig.~\pref{FIG:fig1}{e}. Our ability to impose hard-wall boundaries is a relatively unique feature not usually found in atomic physics experiments, where gases of cold atoms are commonly trapped in smoothly varying laser potentials. We expect that this ability to engineer system edges, dislocations, and defects will be of particular utility in the study of boundary modes in topological systems~\cite{SSH-1979,vonKlitz-NewMethod,HasanRMP-Colloq}, similar to recent studies based on the use of internal hyperfine states as sites in an effective ``artificial dimension''~\cite{Celi-ArtificialDim,Fallani-chiral-2015,Stuhl-Edge-2015}.

\begin{figure}[bt]
\capstart
\vspace{2pt}
\includegraphics[width=\columnwidth]{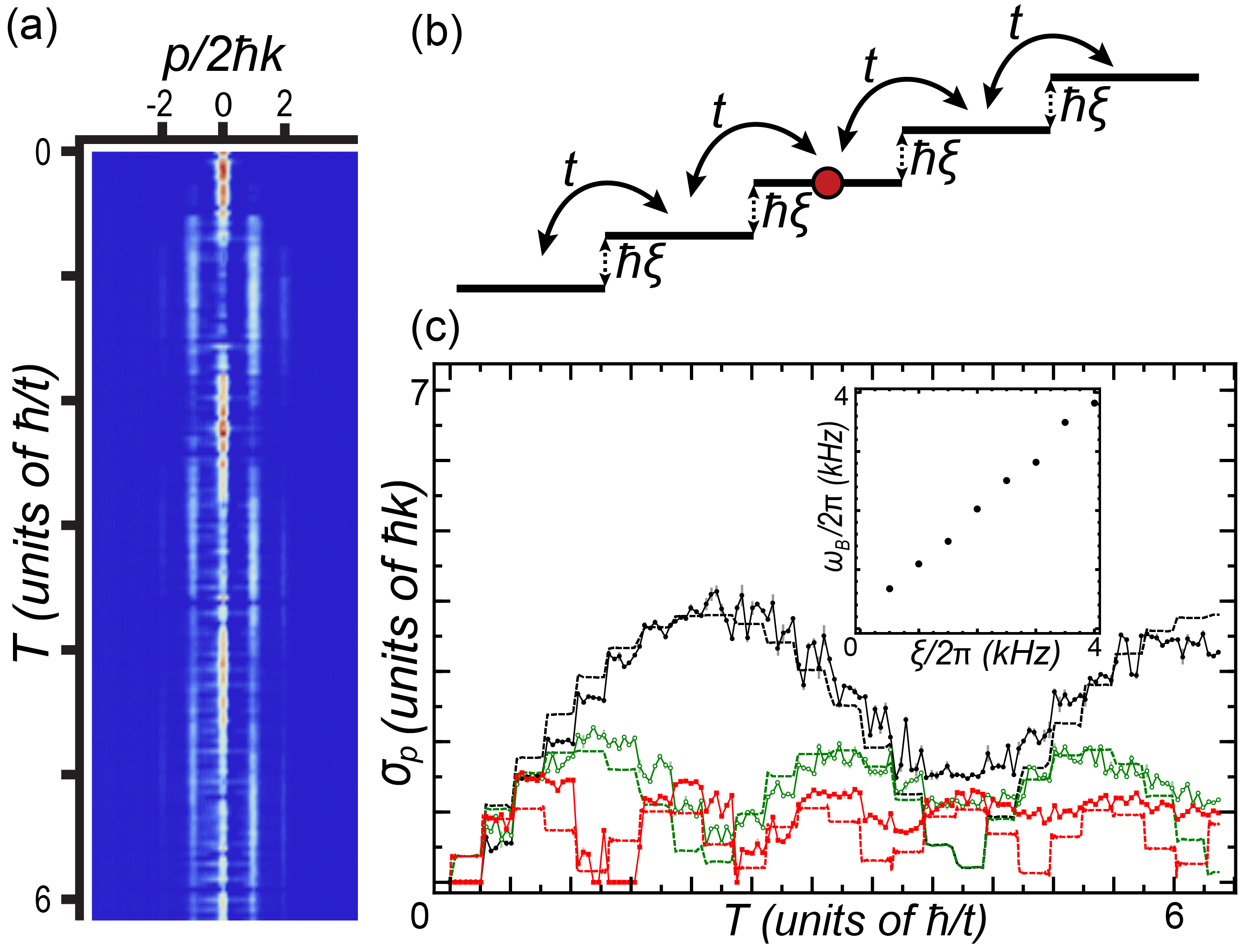}
\caption{Bloch oscillations in a linear potential gradient.
(a)~Integrated momentum spectra vs. evolution time, with uniform $t_n = t \approx 0.33 E_R$ with $n \in \{-10,9\}$ for $\hbar\xi/E_R \approx 1$.
(b)~By detuning all two-photon transition frequencies from resonance by an equal amount $\hbar \xi$, we simulate dynamics of particle transport in a linear potential of the form $\varepsilon_{n+1} = \varepsilon_n + \hbar\xi$.
(c)~Standard deviation of the population distributions vs. evolution time for detunings of $\hbar\xi/E_R \approx 0.5, 1, 1.75$ (black circles, open green circles, red squares), along with theoretical predictions (dashed lines). The inset shows the effective Bloch frequency vs. detuning, as determined from an oscillatory fit, which is well controlled by the two-photon detunings $\xi$.
}
    \label{FIG:fig2}
\end{figure}

\emph{Transport on a tilted lattice.} Having demonstrated the ability to study coherent dynamics in a hopping model with local control of nearest-neighbor coupling strengths, we now explore our ability to control the potential landscape of individual site energies $\varepsilon_n$. As a simple example, we study the spreading of localized wave packets on a 21 ``site'' lattice with uniform tunneling energies $t$ and a linear potential of site energies $\varepsilon_{n+1} = \varepsilon_n + \hbar \xi$. This tilted potential of $\varepsilon_n$ values is achieved through a uniform frequency detuning $\xi$ of all two-photon drives from their respective Bragg resonances, and mimics a uniform force on the particles, analogous to the case of electrons in a uniform electric field.

Nearly a century ago, the motion of particles in a periodic potential under the influence of uniform force was predicted to result in oscillatory motion, so-called Bloch oscillations~\cite{BO-Bloch,BO-Zener}, rather than uniform drift or free acceleration. Over the past several decades, this coherent wave effect has been experimentally studied in electronic systems~\cite{BO-SolidState}, with cold atoms in optical lattices~\cite{BenDahan-Bloch}, in optics~\cite{BO-photon}, and even in the rotational excitations of N$_2$ molecules~\cite{BO-N2}. The absence of dissipation in our system prohibits transport of our initially localized wavepackets in the tilted potential, so instead we expect periodic spreading and refocusing in momentum-space, as recently observed using cold atom microscopy~\cite{Preiss13032015}.

Figure~\ref{FIG:fig2} summarizes our experimental observations of Bloch oscillations, based on the controlled engineering of the potential landscape of site energies $\varepsilon_n$. In Fig.~\pref{FIG:fig2}{a}, we show the momentum-space dynamics (time evolution of the integrated TOF images) for the case $\hbar\xi \approx 1E_R$ (with uniform $t_n = t \approx 0.33E_R$ and $n\in \{-10,9\}$). As expected, we observe the absence of acceleration or dc transport of the atomic populations in the momentum-space lattice. The atomic populations instead undergo periodic cycles of delocalization and refocusing at the original position, with a characteristic Bloch frequency given simply by $\omega_B = \xi$. Figure~\pref{FIG:fig2}{b} shows a diagram of the energy landscape to which the atoms are exposed, with site energy offset given by $\hbar\xi$, and uniform tunneling $t$. Figure~\pref{FIG:fig2}{c} shows $\sigma_p$, the standard deviation of the momentum operator, for several different frequency detunings $\xi$, as a function of evolution time. The inset shows the linear relation between observed Bloch frequency, $\omega_B$, and detuning, $\xi$. While this simple example of a uniform potential gradient has been studied in a number of physical systems, our ability to construct arbitrary site energies $\varepsilon_n$ with local and time-dependent control opens up new prospects in the study of disordered and topological systems.

\emph{Rotary spin echo by control of tunneling phases.} In addition to our ability to control the tunneling amplitudes $t_n$ and site energies $\varepsilon_n$, we can directly control the tunneling phases $\varphi_n$ in a local and time-dependent fashion through our control of the individual relative phases of the optical fields that drive the two-photon Bragg transitions. In higher-dimensional systems ($d \geq 2$) or on multiply-connected lattices (i.e. with higher-order couplings beyond just nearest neighbors), this ability to engineer tunneling phases in an inhomogeneous fashion allows for a direct construction of artificial U(1) gauge fields. This can be used, for example, to study physics associated with the integer Hall effect~\cite{vonKlitz-NewMethod,vonKlitzing-IQHE-1986} or random-flux models~\cite{Lee-Fisher-RandomFlux-1981}. However, in one-dimensional systems with purely nearest-neighbor couplings, any static pattern of inhomogeneous tunneling phases (i.e., any static gauge field) is of no consequence with respect to either equilibrium density distributions or site occupation dynamics. This results from the fact that these phases can simply be ``gauged away'' via local transformations.

Still, we are able to demonstrate our ability to control the tunneling phases through their dynamical variation. Specifically, we demonstrate the reversal of momentum-space dynamics by periodic phase inversions of the form $\varphi \rightarrow \varphi + \pi$. For a dispersive lattice with uniform tunneling amplitudes $t$, this phase inversion can be thought of as band inversion $t \rightarrow -t$, leading to a complete reversal of dynamics such as in the case of light propagation in negative index materials. More directly, this can be thought of as the higher-spin version of a rotary spin echo sequence~\cite{RotaryEcho}.

\begin{figure}[tb]
\capstart
\vspace{2pt}
\includegraphics[width=\columnwidth]{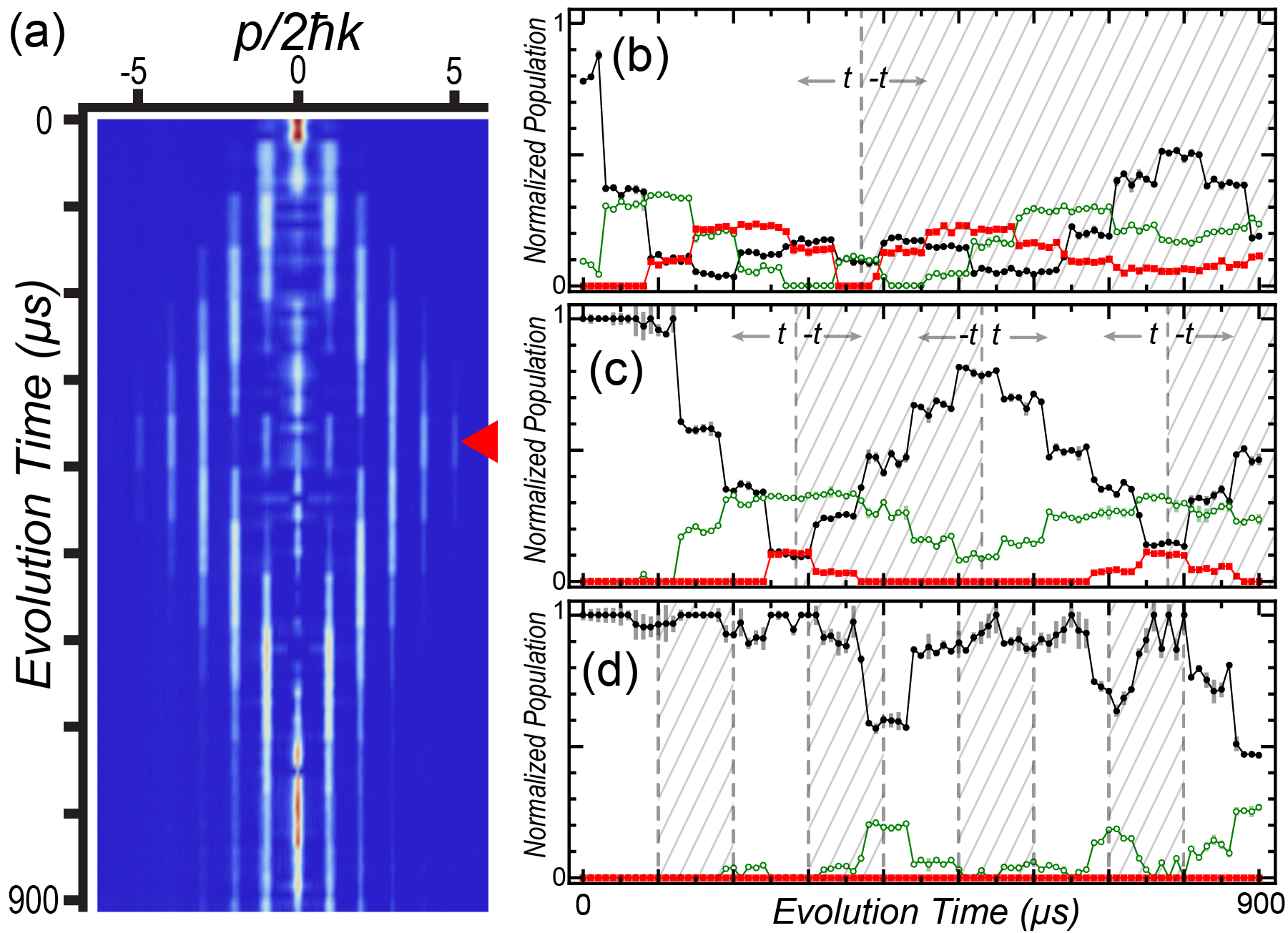}
\caption{Reversal of dynamics through temporal switching of the tunneling phase.
(a)~Integrated momentum spectra vs. evolution time, with the uniform $|t_n| = t \approx 0.3 E_R$ where $n\in\{-10,9\}$, with a phase reversal, $t \rightarrow -t$, at $\sim325~\mathrm{\mu s}$ (indicated by the red marker).
(b-d)~Normalized populations of the zeroth, first, and second (black circles, open green circles, red squares) momentum orders vs. evolution time. As indicated by the dashed vertical lines and shaded regions, the phase is switched one time, three times, and eight times for panels (b), (c), and (d), respectively.
}
\label{FIG:fig3}
\end{figure}

Figure~\ref{FIG:fig3} summarizes our dynamical control over uniform tunneling phases. Figure~\pref{FIG:fig3}{a} shows the full evolution of the momentum spectra, with a reversal of evolution after phase inversion at $325~\mathrm{\mu s}$. The data shown has uniform $t_n = t \approx 0.3 E_R$ with $n\in\{-10,9\}$. The dynamics of momentum mode populations are shown in Fig.~\pref{FIG:fig3}{b}-\pref{FIG:fig3}{d} for several different rates of phase inversion. We can see that for very fast rates of phase inversion (with respect to the tunneling rate $t/\hbar$), transport is inhibited and population remains largely in the central momentum order. While this capability of directly engineering tunneling phases is of somewhat limited utility in one-dimensional models with purely nearest-neighbor tunneling, we expect it to be a powerful tool when extended to higher-dimensional systems or with the inclusion of next-nearest-neighbor tunneling through higher-order Bragg transitions~\cite{Kozuma-Bragg}.

The absence of a perfect reversal of dynamics in Fig.~\ref{FIG:fig3} is likely a consequence of the main practical limitation expected for this experimental scheme -- the loss of spatial mode matching between the different momentum ``sites'' (states). As our trapped sample of atoms initially have some finite coherence length, spatial separation between the differing momentum states $|\psi_n\rangle$ will lead to the loss of coherent momentum-space ``tunneling dynamics'' as driven by two-photon Bragg transitions. We expect that this technical limitation, which may be largely mitigated by working with extended samples of atoms, at the moment presents the greatest source of decoherence in the presented studies.

\emph{Outlook.} We have explored a scheme~\cite{Gadway-KSPACE} for the generation of nearly arbitrary single-particle Hamiltonians for the study of lattice transport phenomena. We demonstrated experimental control over nearest-neighbor tunneling strengths, tunneling phases, and site energies, with the ability to exact local and time-dependent control. We expect that this simple scheme, which also allows for local detection capabilities, will open up many prospects in the experimental study of topological and disordered one-dimensional systems. Additionally, we expect that the study of myriad phenomena will be enabled by straightforward extensions of this technique to larger system sizes, higher dimensions, the inclusion of longer-range hopping, and the use of spin-changing Raman transitions for the study of artificial U(2) gauge fields. Furthermore, the nonlinear and long-ranged (in momentum space, i.e., local in real space) interactions of our atomic matter waves may possibly be harnessed for future studies of interacting topological matter~\cite{2015Meier-footnote-SuppMat}.

\bibliographystyle{apsrev4-1}

\end{document}


\title{Supplemental material for ``Atom-optics simulator of lattice transport phenomena''}
\author{Eric J. Meier}
\author{Fangzhao Alex An}
\author{Bryce Gadway}
\email{bgadway@illinois.edu}
\affiliation{Department of Physics, University of Illinois at Urbana-Champaign, Urbana, Illinois 61801-3080, USA}
\date{\today}


\maketitle
\renewcommand\thefigure{S\arabic{figure}}
\section{Details of experimental scheme}

\begin{figure}[b]
\centering
\capstart
\includegraphics[width=\columnwidth]{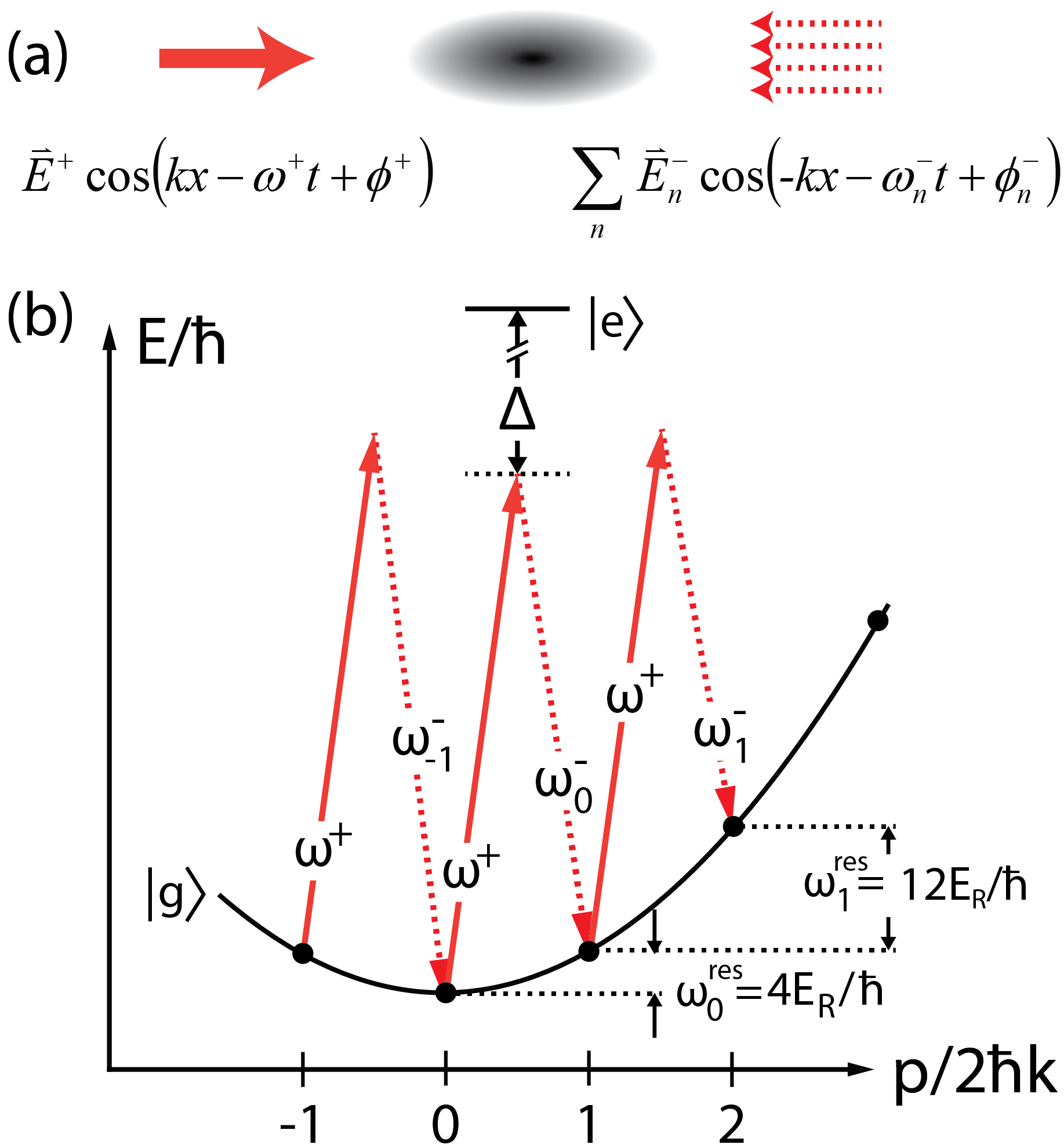}
\caption{Scheme for studying controlled momentum-space transport.
(a)~Cold atoms are driven by counter-propagating laser fields, one made up of several frequency components, controlled in phase, frequency, and amplitude.
(b)~Energy-momentum dispersion. All lasers are far-detuned by $\Delta$ from the ground ($|g\rangle$) to excited ($|e \rangle$) state transition. Stimulated two-photon Bragg transitions coherently couple plane-wave momentum states separated by two photon momenta ($2 \hbar k$). The free particle dispersion defines a unique two-photon Bragg resonance condition $\hbar\omega_n^{\text{res}} = (2n+1)4 E_R$ for each link between neighboring states, with each frequency component of the multi-frequency field addressing a unique link.
}
    \label{FIG:fig1}
\end{figure}

In the main text, we describe an atom optics-based scheme by which we are able to create designer Hamiltonians for the simulation of lattice transport phenomena. This scheme was first introduced in Ref.~\cite{Gadway-KSPACE}, and we reintroduce it with further relevant description in this Supplement. As depicted in Fig.~\ref{FIG:fig1}, the scheme uses stimulated Bragg transitions to control the coherent couplings between a large number of plane-wave momentum states of a Bose--Einstein condensate, mimicking coherent particle tunneling between sites in a lattice array.

The lattice laser wave vector $k$ ($k = 2\pi/\lambda$ with laser wavelength $\lambda$) determines a discrete set of states $\psi_n$ with momenta $p_n = 2n\hbar k$ that may be populated from an at-rest condensate through the stimulated exchange of photons between the two counter-propagating laser fields. As shown in Fig.~\pref{FIG:fig1}{b}, the quadratic energy-momentum dispersion of the massive atoms defines a unique energy difference and Bragg transition frequency $\omega^{\text{res}}_n = (2n+1)4 E_R/\hbar$ (with $E_R = \hbar^2 k^2 / 2M$ the recoil energy and $M$ the mass of the atoms) for each pair of neighboring states $\psi_n$ and $\psi_{n+1}$. Multiple two-photon Bragg transitions are simultaneously driven by writing multiple frequency tones - controlled in amplitude, frequency, and phase - onto one of two counter-propagating and interfering laser fields. This results in spectrally resolved control over all system parameters at the individual link level.

Formally, after adiabatic elimination of the atomic excited state, the momentum-space evolution of the ground-state atoms is governed by the Hamiltonian $\hat{H} = \hat{H}_0 + \hat{V}(t)$. Here, $\hat{H}_0 = \sum_n E_n |\psi_n \rangle \langle \psi_n |$ describes the kinetic energies $E_n = n^2 4 E_R$ of the $\psi_n$ states and $\hat{V}(t) = \sum_n (\chi (t) |\psi_{n+1} \rangle \langle \psi_n | + \chi^* (t) |\psi_{n} \rangle \langle \psi_{n+1} |)$ describes the interaction of ground-state atoms with the counter-propagating far-off-resonant laser fields. The common (for all $n$) off-diagonal coupling constant is defined by $\chi(t) = \sum_j \hbar\tilde{\Omega}_j e^{i \tilde{\phi}_j} e^{-i \tilde{\omega}_j t}$, and relates to changes of momenta by $+2\hbar k$ via virtual absorption of a photon from the right-traveling beam and stimulated emission into one of the left-traveling fields with indices $j$. Here, the two-photon coupling strengths are $\tilde{\Omega}_j = \Omega^-_j \Omega^+ / 2\Delta$ in terms of the single-photon detuning $\Delta$ from atomic resonance (ground $|g\rangle$ to excited $|e\rangle$ state transition) and the single-photon Rabi couplings (assumed to be real-valued), with $\Omega^+$ relating to the right-traveling field and $\Omega^-_j$ to the $j^{\mathrm{th}}$ frequency component of the left-traveling field. Similarly, the two-photon phase shift is related to the phase difference between the interfering fields as $\tilde{\phi}_j = \phi^+ - \phi^{-}_j$, and the two-photon frequency is given by the frequency difference between the interfering fields as $\tilde{\omega}_j = \omega^{-}_j - \omega^+$. We can move to the interaction picture to remove the diagonal kinetic energy terms, leaving only $\hat{H}^{\mathrm{int}}=\hat{V}_I(t)=e^{i\hat{H}_0t/\hbar}\hat{V}(t)e^{-i\hat{H}_0t/\hbar}$, with $\hat{V}_I(t)=\sum_n(\tilde{\chi}(t)|\psi_{n+1}\rangle\langle\psi_n|+\tilde{\chi}^*(t)|\psi_{n}\rangle\langle\psi_{n+1}|)$ and $\tilde{\chi}(t)=\chi(t)e^{i(E_{n+1}-E_n)t/\hbar}=\chi(t)e^{i\omega^{\text{res}}_nt}$.

We achieve our goal of uniquely controlling all nearest-neighbor couplings at the single-link level by associating (up to controlled detunings $\xi_j$) each frequency component with a unique Bragg resonance. Specifically, we define $\tilde{\omega}_j = \omega^{\text{res}}_j - \xi_j$, such that $\chi(t) = \sum_j \hbar\tilde{\Omega}_j e^{i \tilde{\phi}_j} e^{i[(\omega^{\text{res}}_n - \omega^{\text{res}}_j) + \xi_j] t}$. In the weak-driving limit satisfying $\hbar \tilde{\Omega}_j \ll 8 E_R \ \forall \ j$ (also restricting $\hbar |\xi_j| \ll 8 E_R \ \forall \ j$), each frequency component contributes to the coupling of only two particular momentum states $\psi_n$ and $\psi_{n+1}$. Ignoring all rapidly oscillating terms, we arrive at an approximate description with controlled, weakly time-dependent nearest-neighbor couplings $\langle n+1 | \hat{V}_I(t) | n \rangle \approx \hbar\tilde{\Omega}_n e^{i \tilde{\phi}_n} e^{i\xi_n t}$. Finally, we can reabsorb the weak time-dependence due to the $\xi_n$ terms as diagonal site energies $\varepsilon_n$ (with $\hbar\xi_n = \varepsilon_{n+1} - \varepsilon_n$) through a simple redefinition of the state vectors as $|\tilde{\psi}_n \rangle = |\psi_n \rangle e^{i\varepsilon_n t/\hbar}$. This brings us in final form to a highly tunable single-particle Hamiltonian
\begin{equation}
\hat{H}_{\text{eff}} \approx \sum_n \varepsilon_n |\tilde{\psi}_n \rangle \langle \tilde{\psi}_n | + \sum_n t_n ( e^{i\varphi_n} |\tilde{\psi}_{n+1} \rangle \langle \tilde{\psi}_n |  + \mathrm{h.c.}) \ ,
\label{EQ:e0b}
\end{equation}
where arbitrary control over all tunneling amplitudes $t_n \equiv \hbar \tilde{\Omega}_n$, tunneling phases $\varphi_n \equiv \tilde{\phi}_n$, and site energies $\varepsilon_n$ is enabled in a site- and link-dependent fashion through control of the multi-frequency global addressing field.

\section{Experimental details}

\begin{figure}[b]
\centering
\capstart
\includegraphics[width=\columnwidth]{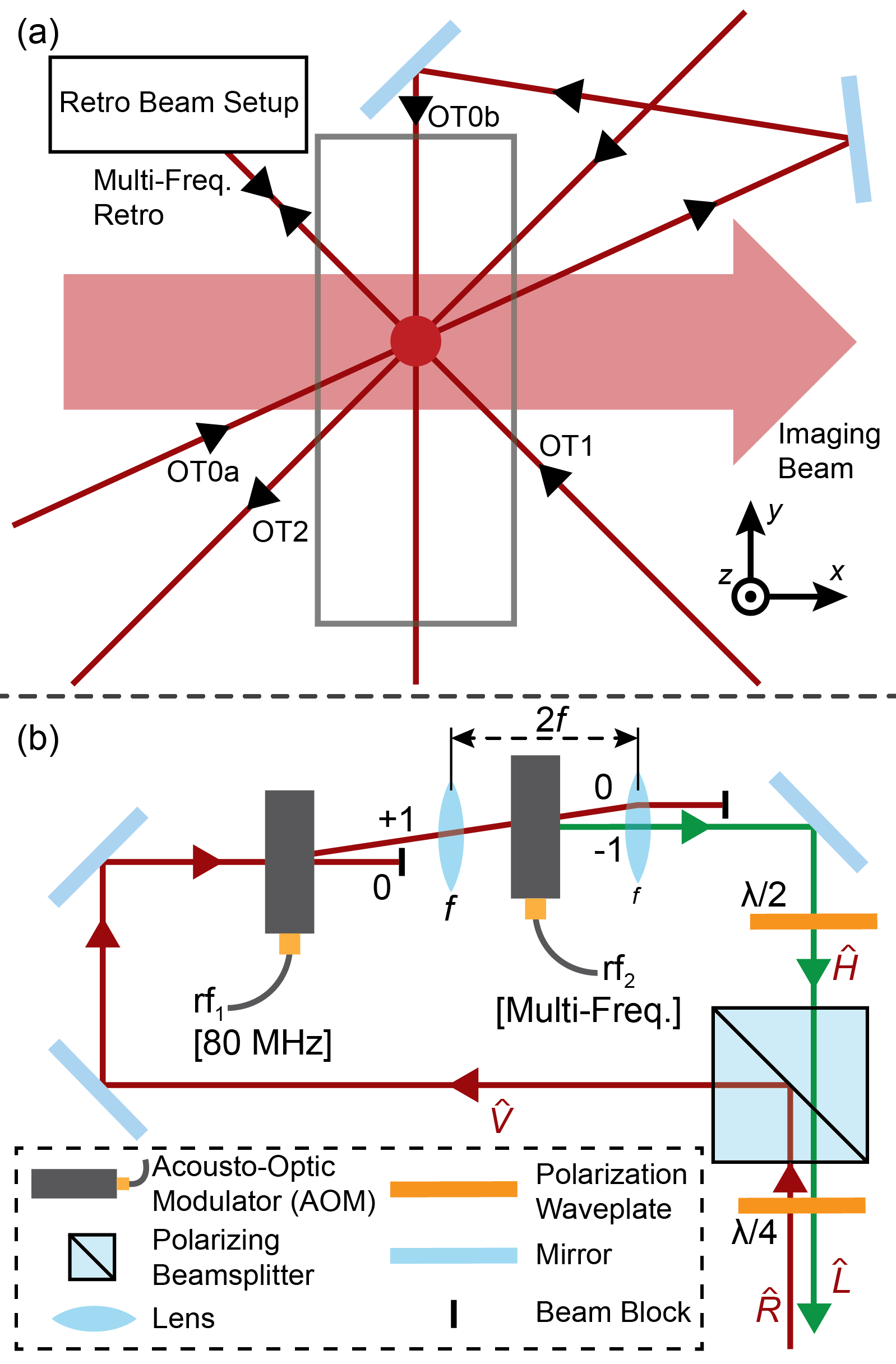}
\caption{Experimental setup.
(a)~Schematic of optical trapping beams and imaging beam. The gray box represents the vacuum chamber, and the red ball represents the atoms. The lattice is created by the interference of OT1 with the mutli-frequency beam.
(b)~Retro beam setup. OT1 is passed through two AOMs such that it acquires any frequency difference present in the two radio frequency driving fields and is then directed to counter-propagate with itself. The lenses focus the beam onto the AOM, resulting in greater diffraction efficiency, and collimate the many frequencies which may be generated at the second AOM.
}
    \label{FIG:fig2}
\end{figure}

Our condensate production begins by first loading a few $10^9$ atoms into a three-dimensional magneto-optical trap (MOT) from a two-dimensional MOT~\cite{Dieckmann-2DMOT-1998} over 10~s, aided by red- and blue-detuned pushing beams~\cite{Park-2colorpush}. Following brief stages of magnetic compression and cooling~\cite{PetrichCompress94, TownsendPSD}, an increase of the density through reduced repumping intensity near the trap center~\cite{Ketterle-DarkSPOT, AndersonDarkMOT}, sub-Doppler cooling via optical molasses~\cite{Lett-SubDoppler}, and optical pumping to the $|F,m_F\rangle = |2,2\rangle$ hyperfine Zeeman sublevel, we extinguish all MOT laser light and capture roughly $5 \times 10^6$ atoms in a far-detuned optical dipole trap (ODT). The ODT consists of both large- and small-volume crossed-beam traps~\cite{Grimm-Kraemer-OT}, with two crossed laser beams (OT0a and OT0b) of wavelength $1070~\mathrm{nm}$ and beam waists ($1/e^2$ radii) of $\sim250~\mathrm{\mu m}$ forming the large-volume trap and two beams (OT1 and OT2) with wavelength 1064~nm and beam waists of $\sim80~\mathrm{\mu m}$ forming the small-volume trap. The optical layout of the laser beams used for trapping, lattice-addressing, and imaging is shown in Fig.~\pref{FIG:fig2}{a}. We perform evaporative cooling along the direction of gravity by exponentially reducing the power in all four beams over six seconds, resulting in pure condensates of roughly $10^5$ atoms.

Prior to performing controlled laser-addressing as described above, we perform an adiabatic decompression of the trap stiffness~\cite{Kozuma-Bragg} so that the condensate becomes extended along the direction of lattice-addressing and momentum-transfer. This increases the timescale over which different momentum-space wave packets are spatially overlapped - i.e. the near-field regime in which the momentum states share a common spatial mode - thus preserving coherent momentum-space dynamics~\cite{Gadway-KSPACE}. At the end of the adiabatic decompression procedure, condensates containing roughly $5 \times 10^4$ atoms are confined almost entirely by OT1, such that they are spatially extended along one direction. Weak additional confinement is provided by beams OT0a and OT0b, resulting in trapping frequencies of $\omega_{\{1,2,3\}} \approx 2\pi \times \{130,10,130\}~\mathrm{Hz}$.

The trapping laser beam OT1 serves as one of the two interfering laser beams that drives the momentum-space dynamics. The counter-propagating field, composed of multiple, evenly spaced frequency components in the same spatial mode, is derived by passing OT1 through two acousto-optic modulators (AOMs) in a single-pass configuration with diffraction in the $+1^{\mathrm{st}}$ and $-1^{\mathrm{st}}$ diffraction orders, respectively. The diffracted laser beam now has the multiple, evenly spaced frequency components written onto it through the controlled radio frequency driving of the second AOM. This beam is then aligned to counter-propagate with the OT1 trapping beam. The optical setup that produces this counter-propagating laser field, including the relevant optical lenses and polarization optics, is shown in Fig.~\pref{FIG:fig2}{b}. The interaction of the atoms with the interfering laser beams results in driven momentum-space dynamics as described above, characterized by changes in velocity of $\pm 2 v_R \approx \pm 8.6~\mathrm{\mu m/ms}$ and a recoil energy $E_R \approx h \times 2.03~\mathrm{kHz}$. After a chosen evolution time, during which the momentum-space dynamics are governed by the effective Hamiltonian of Eq.~\eqpref{EQ:e0b}, all laser fields are turned off and the atoms are allowed to freely fall and expand in time of flight (TOF). An absorption image is taken after 18~ms of TOF, revealing the populations in the different momentum orders (states $\psi_n$). The imaging beam propagates at 45 degrees with respect to the direction of momentum transfer, such that the separate momentum orders can be uniquely identified.

\section{Accuracy of Parameter Control}

The three parameters over which we demonstrate full control are the tunneling amplitudes, the tunneling phases, and the site energies. The tunneling phases and the site energies are essentially exact, as they are controlled through direct digital radio frequency engineering. The tunneling amplitudes, however, are subject to uncertainty due to beam alignment and therefore must be measured.

\begin{figure}[t]
\centering
\capstart
\includegraphics[width=\columnwidth]{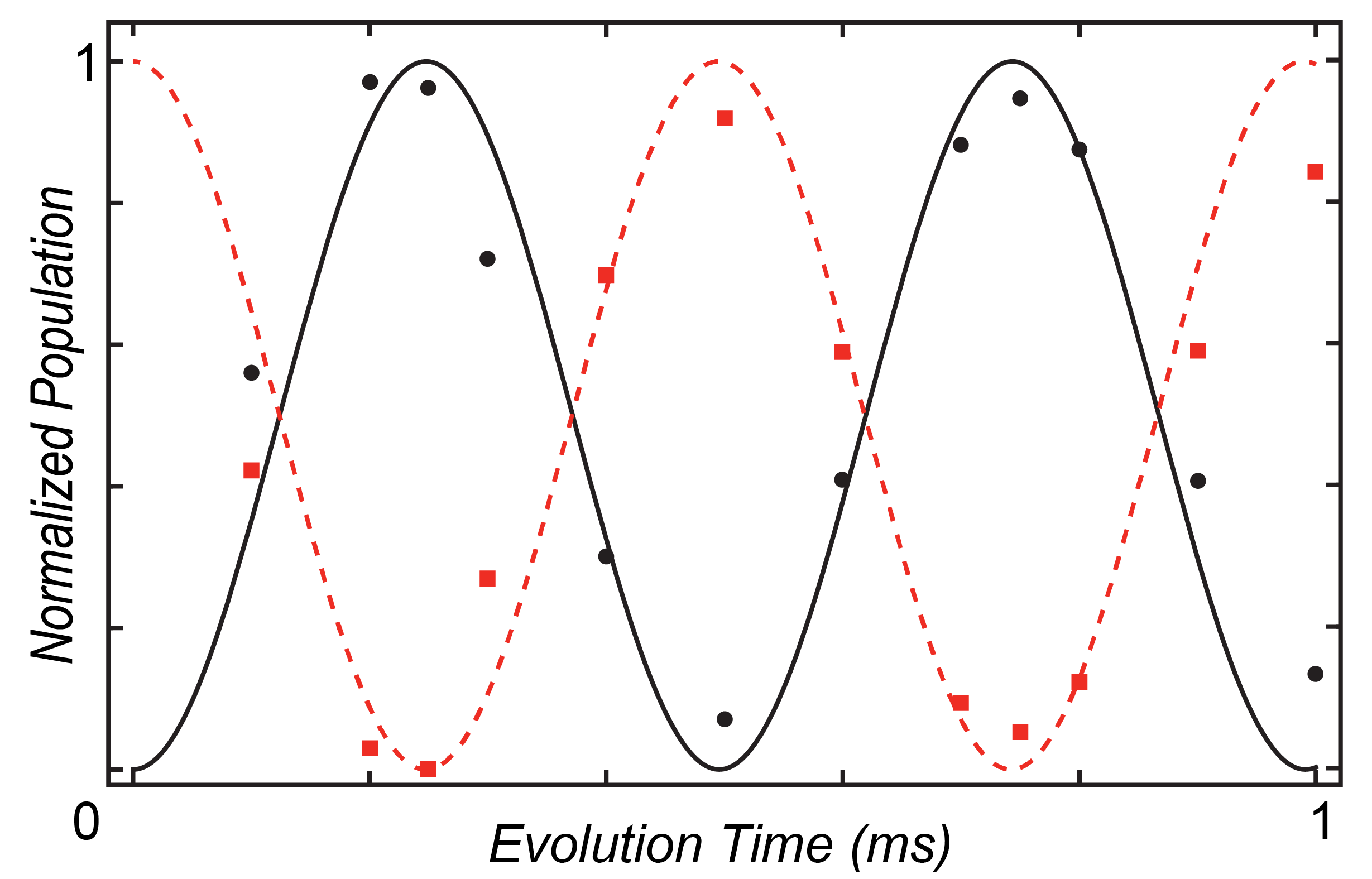}
\caption{Example tunneling amplitude calibration curve. Normalized population of the zeroth (black circles) and first (red squares) momentum orders as a function of evolution time. The solid black line represents a fit to the zeroth order population with the function $\sin^2(\pi f T)$ and the dashed red line represents a fit to the first order with the function $\cos^2(\pi f T)$, where $f$ is the free parameter and $T$ is the evolution time in both cases. This calibration results in uncertainties in the tunneling amplitude of a few percent ($1.3\%$ for the data pictured).
}
    \label{FIG:fig3}
\end{figure}

We calibrate the tunneling amplitudes by creating a multi-frequency counter-propagating beam that links only the $0 \hbar k$ and $+2 \hbar k$ momentum orders with many frequency couplings far from relevant two-photon Bragg resonances for radio frequency power control. We then observe population undergo Rabi oscillations between these two connected orders. Fig.~\ref{FIG:fig3} shows a typical set of curves for this process. The points are the normalized population of the zeroth (black circles) and first (red squares) momentum orders versus evolution time. The accompanying lines are fits to the zeroth and first order population of the form $\sin^2(\pi f T)$ and $\cos^2(\pi f T)$, where $f$ is the free parameter and $T$ is the evolution time. The Rabi rate, $f$, is related to the tunneling amplitude by $t=(1/4) h f$, where $h$ is the Planck constant. These fits give a typical uncertainty in the tunneling amplitude of a few percent ($1.3\%$ in the case of the data shown in Fig. \ref{FIG:fig3}).

\section{Image and Data Analysis and Influence of Atomic Interactions}

We perform absorption imaging of our $\ket{F,m_F} = \ket{2,2}$ atoms, using a near-resonant laser beam as shown in Fig.~\pref{FIG:fig2}{a}. The resulting optical depth images of our atomic condensates, such as those shown in Fig.~1(a) of the main text, are then analyzed to determine details of the momentum-space distributions. We numerically integrate the images along the z-axis to determine the one-dimensional momentum profiles. Because of the sensitivity of our measurements to small amounts of noise at high momentum values, we artificially reduce our image noise by first filtering out pixel values below a threshold value prior to integration. This threshold level is determined from our characteristic image noise, by making a Gaussian fit to the distribution of pixel counts in image areas with no atomic signal.

This momentum profile is fit with a summation of Gaussian functions, one for each possible momentum order. Each Gaussian is constrained to be at a position given by the constant $2\hbar k$ spacing of the momentum peaks relative to the position of the zero momentum condensate. In addition, there is a group of non-condensate atoms centered about zero momentum. These thermal atoms are fit by a separate Gaussian with parameterized height and width for each image. This fitting procedure results in an amplitude and fit error for the number of atoms in each momentum order. To calculate the average population for each image, many repeated measurements are weighted according to their fit errors and averaged. These populations are then normalized for each image and any value below $\approx 1\%$ is set to zero as the fits are dominated by noise at this level. The remaining populations are then renormalized to give the final population values.

We note that some influence of atomic interactions is readily observed in the analyzed data. For situations where most of the atomic population is restricted to only a few low momentum orders, $s$-wave collisions lead to appreciable effects during TOF as the wave packets of the different momentum orders move through one another. The collision of these condensates leads to the pairwise scattering of colliding atoms into a shell of momentum states allowed by momentum and energy conservation ($s$-wave halos). This elastic loss of population from the discrete set of plane-wave momentum states that we consider (and measure) to other momenta has some influence on the extracted fractional momentum-state populations at short evolution times. Specifically, as the absolute number of atoms in two colliding condensates will be reduced by the same amount due to $s$-wave collisions (occurring predominantly during TOF), the measured fractional population in one of these states will be reduced or enhanced if it initially had a relative deficiency or excess of atoms. At short evolution times, when there is very little population in any order other than zero, this effect can cause the overestimation of the fraction of population in the zeroth order. We believe these elastic ``losses'' account for the discrepancies between the experimental data and theoretical predictions at short evolution times (cf. Fig.~1(c) of the main text).

We remark that in addition to these deleterious effects of atomic interactions, the nonlinear interactions between our atomic matter waves may in the future be harnessed in a useful way for these simulation studies. In particular, the short-ranged interactions of our atoms in real space relate to effectively long-ranged interactions between atoms in the different plane wave momentum orders. Naively we may expect the atomic interactions, which at low collision energies can be approximately modeled by a zero-range contact interaction with an $s$-wave scattering length $a$, will lead to an infinite-range all-to-all interaction in momentum space that has no appreciable influence of the atomic dynamics. However, the collisional cross section $\sigma (2 n k)$ for atoms colliding with a relative momentum of $2n\hbar k$ will more generally be reduced as roughly $\sigma (2nk) = 8 \pi a^2 / (1 + 4 n^2 k^2 a^2)$~\cite{Dalibard98collisionaldynamics}. For the large number of momentum orders populated in our experiments, this correction to the low energy scattering of the different momentum orders will lead to a long-ranged interaction in momentum space, the effective range of which may be controlled through either the atomic scattering properties or the magnitude of lattice-induced momentum transfer. By reducing the strength of the single-particle tunneling terms with respect to nonlinear contributions to the wave packet energies, the influence of these interactions may be probed in future experiments.

\bibliographystyle{apsrev4-1}